# Impact of heat waves and cold spells on cause-specific mortality in the city of São Paulo, Brazil

Sara Lopes de Moraes[ab*], Ricardo Almendra[c], Ligia Vizeu Barrozo[ab]

a Department of Geography, School of Philosophy, Literature and Human Sciences of the University of São Paulo, São Paulo, SP, Brazil
b Institute of Advanced Studies, of the University of São Paulo, São Paulo, SP, Brazil.
E-mails: saraldmoraes@gmail.com; lija@usp.br
c Centre of Studies on Geography and Spatial Planning (CEGOT); Department of Geography and Tourism, University of Coimbra, Coimbra, Portugal. E-mail: ricardoalmendra85@gmail.com

*Corresponding author. Department of Geography, School of Philosophy, Literature and Human Sciences, University of São Paulo, São Paulo, SP, Brazil
E-mail address: saraldmoraes@gmail.com (S.L. Moraes)

**Abstract**

The impact of heat waves and cold spells on mortality has become a major public health problem worldwide, especially among older adults living in low- to middle-income countries. This study aimed to investigate the effects of heat waves and cold spells under different definitions on cause-specific mortality among people aged ≥65 years in São Paulo from 2006 to 2015. A quasi-Poisson generalized linear model with a distributed lag model was used to investigate the association between cause-specific mortality and extreme air temperature events. To evaluate the effects of the intensity under different durations, we considered 12 heat wave and nine cold spell definitions. Our results showed an increase in cause-specific deaths related to heat waves and cold spells under several definitions. The highest risk of death related to heat waves was identified mostly at higher temperature thresholds with longer events. We verified that men were more vulnerable to die from an ischemic stroke on heat waves and cold spells days than women, while women presented a higher risk of dying from ischemic heart diseases during cold spells and tended to have a higher risk of chronic obstructive pulmonary disease than men. Identification of heat wave- and cold spell-related mortality is important for the development and promotion of public health measures.

**Keywords:** heat wave; cold spell; mortality; extreme air temperature event; elderly

**Abbreviations:** ETE: Extreme air temperature events; CI: confidence interval; COPD: chronic obstructive pulmonary disease; CVD: cardiovascular disease; df: degrees of freedom; ICD-10: International Classification of Diseases, 10$^{th}$ Revision ns: natural cubic spline; $PM_{10}$: particulate matter with an aerodynamic diameter of <10 μm; RH: relative humidity; RR: relative risk; São Paulo State System for Data Analysis Foundation: SEADE; T: daily mean air temperature.

# 1. Introduction

Extreme weather events, such as heat waves, cold spells, and droughts, have always occurred worldwide. However, in the past few decades, the occurrence of these extreme events has increased owing to the current climate change (IPCC, 2014). In the near future, extreme air temperature events (ETEs), particularly events related to heat will be more frequent, longer, and intense (Meehl and Tebaldi, 2004; Perkins et al., 2012).

The definitions of ETEs are inconsistent in the literature, and no standard criteria have been established because of the differences in geographical locations, climate variability across regions, and population acclimatization (Robinson, 2001). However, ETEs are generally defined as periods of extremely high or low daily temperatures (mean, maximum, or minimum) outside the normal relative or absolute threshold, which last for consecutive days (Guo et al., 2017; Monteiro et al., 2013; Robinson, 2001). The study of various ETE definitions and different combinations may support the establishment of the best predictor to quantify the impact of ETEs on human health.

Several epidemiologic studies (Chen et al., 2020; Wang et al., 2016; Yang et al., 2019) have reported that the intensity and duration of ETEs influence mortality and morbidity, making it a current public health concern. A recent study found that heat wave-related increase in deaths is projected to be higher in Brazil and other tropical and subtropical areas than in the US and European countries, especially if no mitigation and adaptation strategies are applied to reduce the effects of heat waves on human health (Guo et al., 2018). Although in recent years, most studies have investigated the effects of heat waves and heat events on mortality, the high risk of deaths related to cold events is also expected to remain in some areas (Gasparrini et al., 2015, 2017).

Extremely high and low temperatures can induce substantial physiological stress in the human body. Furthermore, exposure to direct or indirect extreme cold and heat can

4trigger cardiovascular and respiratory symptoms, particularly in older adults, which is the most vulnerable group. Older people present physiological and socioeconomic limitations and are at a higher risk of death from cardiovascular and respiratory diseases (Anderson and Bell, 2009; Chen et al., 2019; Song et al., 2018; Vasconcelos et al., 2013; Wang et al., 2016)

Statistically significant associations between ETEs and the risk of mortality among people aged ≥65 years have been widely reported (Chen et al., 2019; Wang et al., 2016; Yang et al., 2019; Yin et al., 2018). In the summer of 2003, a severe heat wave event caused 70,000 excess deaths across European countries, affecting mainly older adults in France, Portugal, Luxemburg, and Italy (Robine et al., 2008). The excess mortality during extreme ETEs among older individuals occurred not only in Europe but also in Asia, Australia, and the United States over the years (Anderson and Bell, 2009; Tong et al., 2013; Yang et al., 2019). Most recently, in June 2021, an exceptional early hot summer condition occurred in the northwest and western Canada. Approximately 500 people may have died from heat wave events, especially older adults living alone in the greater Vancouver area (Cecco, 2021). In the same week, extremely cold temperatures were recorded for several days in the south and southwest of Brazil (INMET, 2021a).

Nonetheless, there is a lack of knowledge on the association between ETEs and excess mortality among the vulnerable population, particularly in Latin American cities. São Paulo is the most populated city in Latin America, with a significant level of socioeconomic inequalities, environmental disparities, and a large aging population. According to the São Paulo State System for Data Analysis Foundation (SEADE), the number of people aged ≥65 years is expected to be 2.8 million by 2050, representing about 22.9% of the population and an increase of 2% per year. Therefore, this study aimed to investigate the effects of heat waves and cold spells under different definitions on cause-specific mortality among people aged ≥65 years in São Paulo from 2006 to 2015.

## 2. Material and methods

### 2.1. Study area

The city of São Paulo is located in southeast Brazil (Figure 1), with an estimated population of approximately 12.3 million and average population density of 7,398.26 hab/km$^2$ in 2020 (IBGE, 2020). According to the Köppen–Geiger classification, the study area has a humid subtropical climate, characterized by hot and wet summers and dry winters. The average monthly temperatures vary between 16.7 °C and 23.2 °C (INMET, 2021b).

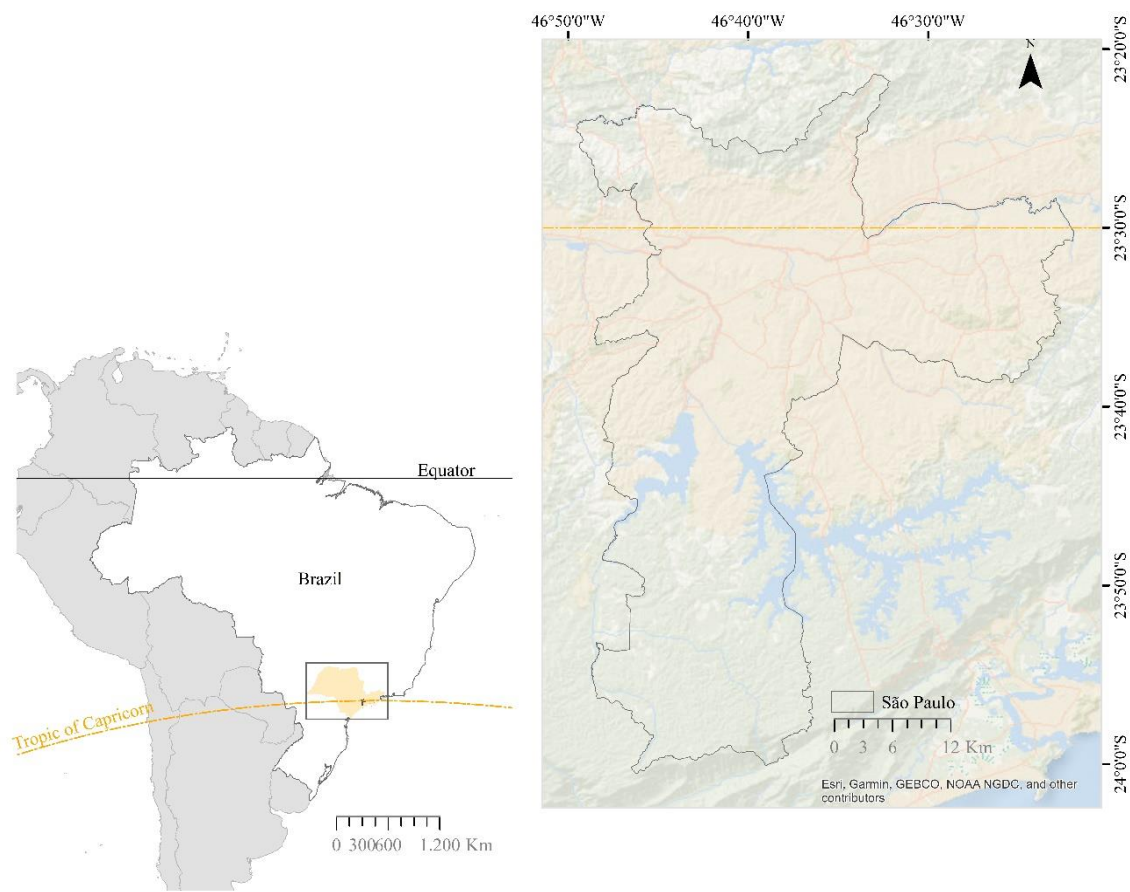

Figure 1: City of São Paulo in Southeast Brazil

*2.2. Data collection*

The daily mortality data of individuals aged ≥65 years from 2006 to 2015 were provided by the Information Improvement Program of the Municipality of São Paulo. For this study, we selected deaths due to cardiovascular diseases (CVD; International Classification of Diseases, 10$^{th}$ Revision [ICD-10]: I00–I99), respiratory diseases (ICD-10: J00–J99), and the following selected specific causes of death within these groups: ischemic heart disease (ICD-10: I20–I25), cerebrovascular diseases (ICD-10: I60–I69), ischemic stroke (ICD10: I63 and I65–I66), hemorrhagic stroke (ICD-10: I60–I62), and chronic obstructive pulmonary disease (COPD; ICD-10: J40–J44 and J47). Additionally, we divided the daily deaths by sex (female and male) for each cause of mortality.

Daily mean air temperature (T; ºC) and relative humidity (RH; %) data for the study period were collected from the meteorological station of the Institute of Astronomy, Geophysics, and Atmospheric Sciences at University of São Paulo.

To adjust the effects of air pollution on the model, data on daily average concentrations of particulate matter with an aerodynamic diameter of <10 μm ($PM_{10}$) were obtained from the Environmental Company of the State of São Paulo.

*2.3. Extreme air temperature event definitions*

The T was assessed to classify the ETEs. The warm period was between September and March and the cold period between April and August. In this study, we analyzed heat wave events according to four relative thresholds (above the 90$^{th}$, 92.5$^{th}$, 95$^{th}$, and 97.5$^{th}$ percentiles) and durations over 2, 3, or 4 consecutive days. Cold spells were defined by three relative thresholds (below the 3$^{rd}$, 5$^{th}$, and 10$^{th}$ percentiles) and over 2, 3, or 4 consecutive days. Therefore, 12 heat wave and nine cold spell definitions were individually assessed.

*2.4. Statistical analyses*

To analyze the association between daily mortality and the different ETE definitions presented, we applied a quasi-Poisson generalized linear model with a distributed linear lag model to capture the lag effects, as shown in previous studies (Chen et al., 2019; Yin et al., 2018; Zhao et al., 2019). We conducted a separate analysis of the heat wave and cold spell models. The general form of the model is as follows:

$$\text{Log}(\mu_t) = \alpha + cb(ETE_t, \text{lag}) + ns(RH_t, df) + ns(PM_{10t}, df) + ns(\text{time}_t, df*10) + ns(Dos_t, df) + \gamma Dow_t + \delta Holiday_t$$

where $\mu_t$ is the daily number of deaths on day $t$ of observation; $\alpha$ is the intercept; ETE is a binary variable that represents the heat wave and cold spell event on day $t$ (1 = heat wave/cold spell days, and 0 = non-heat wave/cold spell days); *cb* is the *crossbasis* function with a linear function and a natural cubic spline function (*ns*) with four degrees of freedom (*df*) for lagged effects on mortality in the heat wave model and 3 df in the cold spell models. Based on several studies (Chen et al., 2019; Guo et al., 2017; Yang et al., 2019), we used 10 lag days for heat waves and 27 lag days for cold spells.

Potential confounders, such as RH and $PM_{10}$, were controlled in the overall ETE model through the ns function. For the heat wave model, the RH and $PM_{10}$ were controlled with 2 df. The models were also adjusted for long-term trends (time) of 2 df per year. Seasonality was controlled using the ns function with 2 df for the day of the season (Dos). Public holidays (Holiday) and the day of the week (Dow) were added to the model as categorical variables, which were in accordance with recent investigations (Guo et al., 2017; Yin et al., 2018).

Furthermore, in the second stage, we included the daily mean temperature in the previous model to verify the added effects of ETE on mortality. The daily mean temperature was controlled using the ns function and 2 df.

We assessed the cumulative relative risk (RR) with its 95% confidence interval (CI) for cold spell days compared to non-cold spell days and for heat wave days compared to non-heat wave days separately for each ETE definition, cause-specific death, and sex.

A sensitivity analysis was conducted to calibrate the model parameters by changing the df for RH, air pollution (2–4), T lag days (heat waves: 0–7 and 0–10 days; cold spells: 0–21 and 0–27 days), and lag days (2–4). We also changed the df for seasonality and long-term trends (1–4). The model parameters were selected in this phase; preference was given to the lower values of the Akaike information criterion for quasi-Poisson regression.

All statistical analyses were performed using R software version 4.0.2. (R Core Team, 2019) with the *dlnm* package (Gasparrini, 2011).

## 3. Results

Table 1 summarizes the descriptive statistics of daily mortality and meteorological and air pollution data. Between 2006 and 2015, there were 151,001 deaths in São Paulo due to CVD and 64,778 deaths due to respiratory diseases. Of these, a total of 56,885 ischemic heart disease deaths were registered. Among them, 38,084 were overall cerebrovascular disease mortalities (ischemic stroke: 11,427; hemorrhagic stroke: 7,718), and 19,148 COPD mortalities.

Table 1: Descriptive statistics of mortality, meteorological, and air pollution data

| Variables | Mean | SD | Min | Median | Max |
|---|---|---|---|---|---|
| Cardiovascular diseases | 41.3 | 8.3 | 18 | 41 | 77 |
| Respiratory diseases | 17.7 | 5.3 | 3 | 17 | 44 |
| Cerebrovascular diseases | 10.4 | 3.4 | 1 | 10 | 26 |
| Ischemic stroke | 3.1 | 2.0 | 0 | 3 | 14 |
| Hemorrhagic stroke | 2.1 | 1.5 | 0 | 2 | 9 |
| Ischemic heart diseases | 15.6 | 4.6 | 3 | 15 | 35 |
| Chronic obstructive pulmonary disease | 5.2 | 2.4 | 0 | 5 | 17 |
| Mean temperature (ºC) | 19.6 | 3.3 | 7.3 | 19.7 | 28.0 |
| Relative humidity (%) | 80.0 | 8.8 | 34.3 | 80.9 | 97.0 |
| $PM_{10}$ (µg/m³) | 35.5 | 17.1 | 7.4 | 31.6 | 132.4 |

SD: Standard deviation; Min: minimum; Max: maximum; $PM_{10}$: particulate matter with an aerodynamic diameter of <10 μm.

During the study period, most of the heat wave events occurred in February, while cold spells events were registered in July. The annual mean number of cold spells on consecutive days for each definition ranged from 2 to 32 days, and that of heat waves on consecutive days for each definition ranged from 2 to 28 days. Detailed information on the descriptive statistics of heat waves and cold spell days for each definition is presented in Table 2.

Table 2: Descriptive statistics of a heat wave and cold spell days for each definition in São Paulo, 2006–2015

| ETE model name | ETE definition | ETE days per year | | | | |
|---|---|---|---|---|---|---|
| | | Mean | SD | Min | Median | Max |
| HW_90P_2d | Heatwave >90th percentile with ≥2 days duration | 28 | 13 | 9 | 25 | 55 |
| HW_90P_3d | Heatwave >90th percentile with ≥3 days duration | 20 | 12 | 3 | 15 | 43 |
| HW_90P_4d | Heatwave >90th percentile with ≥4 days duration | 14 | 12 | 0 | 8 | 34 |
| HW_92.5P_2d | Heatwave >92.5th percentile with ≥2 days duration | 19 | 13 | 4 | 16 | 48 |
| HW_92.5P_3d | Heatwave >92.5th percentile with ≥3 days duration | 12 | 10 | 0 | 9 | 34 |
| HW_92.5P_4d | Heatwave >92.5th percentile with ≥4 days duration | 9 | 9 | 0 | 6 | 28 |
| HW_95P_2d | Heatwave >95th percentile with ≥2 days duration | 12 | 10 | 0 | 8 | 34 |
| HW_95P_3d | Heatwave >95th percentile with ≥3 days duration | 8 | 8 | 0 | 6 | 24 |
| HW_95P_4d | Heatwave >95th percentile with ≥4 days duration | 5 | 7 | 0 | 3 | 21 |
| HW_97.5P_2d | Heatwave >97.5th percentile with ≥2 days duration | 5 | 6 | 0 | 3 | 18 |
| HW_97.5P_3d | Heatwave >97.5th percentile with ≥3 days duration | 3 | 5 | 0 | 0 | 16 |
| HW_97.5P_4d | Heatwave >97.5th percentile with ≥4 days duration | 2 | 5 | 0 | 0 | 16 |
| CS_3P_2d | Cold spell <3rd percentile with ≥2 days duration | 8 | 5 | 0 | 9 | 13 |
| CS_3P_3d | Cold spell <3rd percentile with ≥3 days duration | 4 | 4 | 0 | 5 | 11 |
| CS_3P_4d | Cold spell <3rd percentile with ≥4 days duration | 2 | 3 | 0 | 0 | 7 |
| CS_5P_2d | Cold spell <5rd percentile with ≥2 days duration | 15 | 7 | 2 | 15 | 27 |
| CS_5P_3d | Cold spell <5rd percentile with ≥3 days duration | 9 | 6 | 0 | 9 | 19 |
| CS_5P_4d | Cold spell <5rd percentile with ≥4 days duration | 6 | 5 | 0 | 5 | 13 |
| CS_10P_2d | Cold spell <10rd percentile with ≥2 days duration | 32 | 9 | 12 | 33 | 43 |
| CS_10P_3d | Cold spell <10rd percentile with ≥3 days duration | 24 | 9 | 8 | 22 | 43 |
| CS_10P_4d | Cold spell <10rd percentile with ≥4 days duration | 14 | 7 | 5 | 14 | 25 |

SD: Standard deviation; Min: minimum; Max: maximum

Figure 2 and Supplementary Table S1 illustrate the cumulative RRs of the association between heat waves under 12 definitions and cause-specific mortality at lag 0–10 days. In general, we found that a higher temperature threshold and duration (HW_95P_3d, HW_95P_4d, HW_97.5P_3d, and HW_97.5P_4d) had a higher significant risk, except for hemorrhagic stroke and COPD outcomes. The overall effects (without controlling for daily mean temperature) of heat waves were stronger on mortality due to ischemic stroke, especially in the HW_90P_3d and HW_90P_4d definitions with RR = 1.516 (95% CI: 1.08–2.123) and RR = 1.577 (95% CI: 1.083–2.296), respectively.

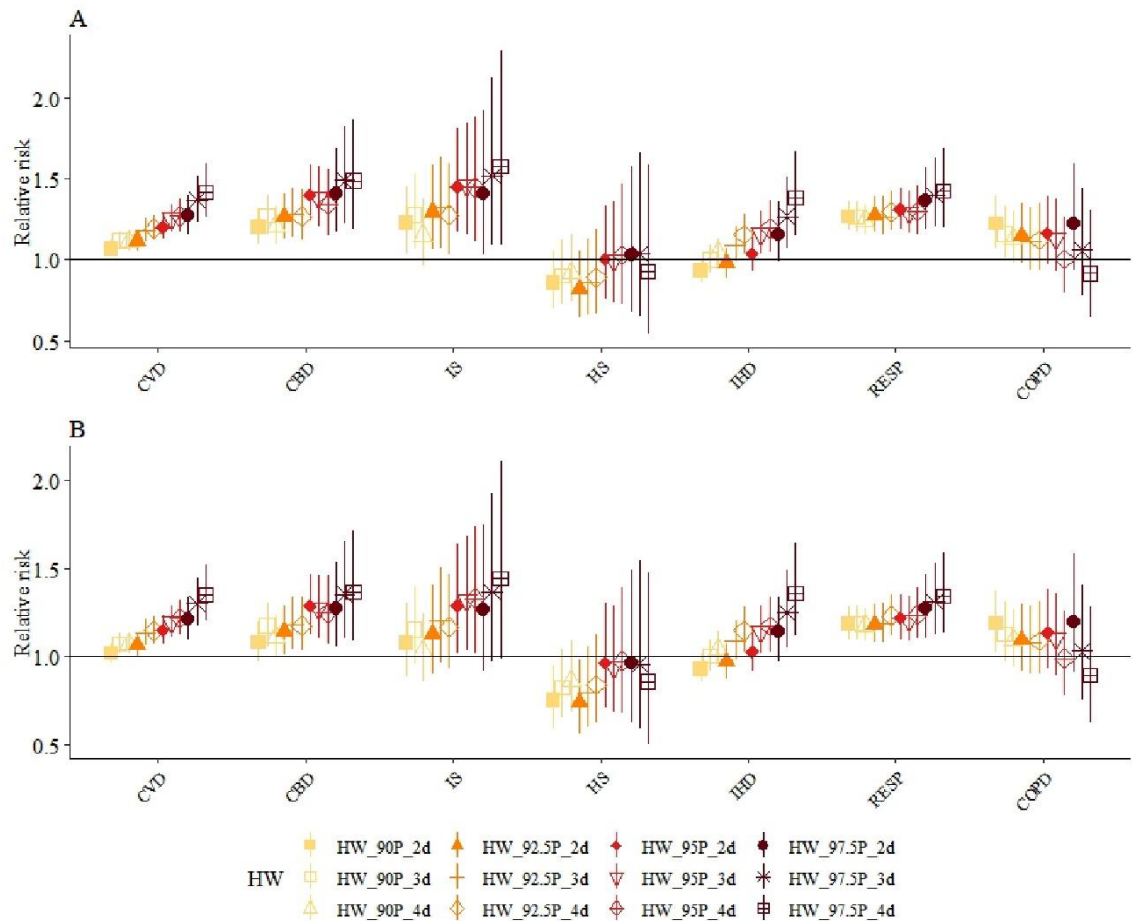

Figure 2: Effects of heat waves under 12 definitions on mortality due to cardiovascular disease (CVD), cerebrovascular disease (CBD), ischemic stroke (IS), hemorrhagic stroke (HS), ischemic heart disease (IHD), respiratory disease (RESP), and chronic obstructive pulmonary disease (COPD) in people aged ≥65 years over lag 0–10 days in São Paulo. (A) Overall effects of heat waves without controlling for daily mean temperature (B) Added effects of heat waves after controlling for daily mean temperature.

Figure 3 displays the cumulative RRs of the impact of cold spells on mortality. The overall cold spell effects were significant in several outcomes and definitions. The effect estimates varied greatly according to the definition and outcome. The RRs of CVD, for instance, varied from 1.216 (95% CI: 1.026–1.442) in the CS_10P_2d definition to 2.484 (95% CI: 1.456–4.239) in the CS_3P_4d definition. In addition, the significant RR values for the cause-specific mortality of cerebrovascular diseases, ischemic stroke, and ischemic heart diseases were more pronounced in some cold spells than during heat waves.

The added effects in the heat waves and cold spell models were also significant for several definitions. The added effects of the heatwave models showed lower RRs compared to the overall effects in all cause-specific mortality. In addition, we observed a higher risk of death on the added effects in comparison with the overall effects for cold spell models, especially for CVD (CS_10P_2d), respiratory diseases (except the 10[th] percentile temperature threshold with ≥2 and ≥3 days), hemorrhagic stroke (CS_10P_4d, CS_5P_2d, CS_3P_3d, and CS_3P_4d), and for all statistically significant RRs of cerebrovascular diseases and ischemic stroke outcomes.

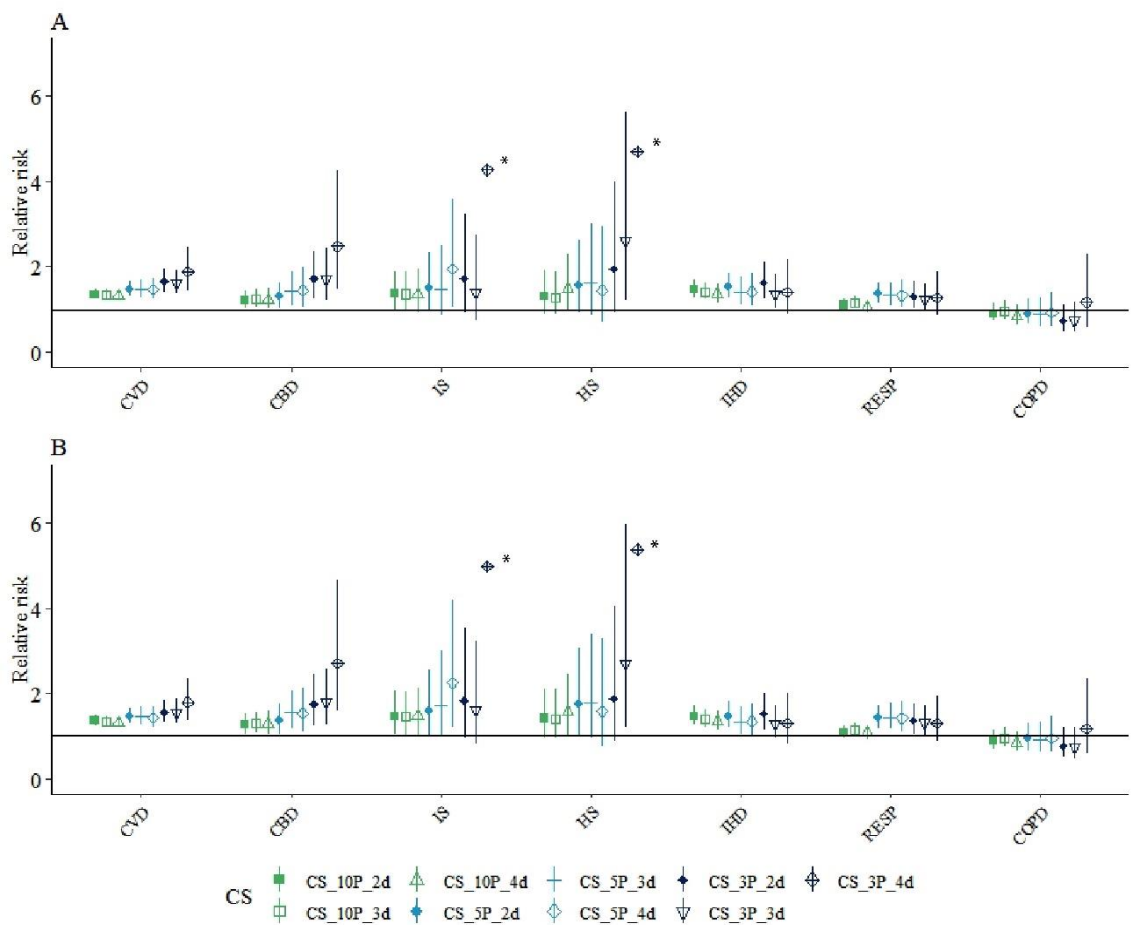

Figure 3: Effects of cold spells under nine definitions on mortality due to cardiovascular disease (CVD), cerebrovascular disease (CBD), ischemic stroke (IS), hemorrhagic stroke (HS), ischemic heart disease (IHD), respiratory disease (RESP), and chronic obstructive pulmonary disease (COPD) in people aged ≥65 years over lag 0–21 days in São Paulo. (A) Overall effects of cold spells without controlling for daily mean temperature and (B) added effects of cold spells after controlling for daily mean temperature.
* The confidence interval is not represented in the graph for better visualization of the other RRs. Confidence intervals are presented in Supplementary 2.

Figure 4 and Supplementary Tables 3 and 4 show the RRs of the association between and the ETE and cause-specific mortality stratified by sex. The risk of death during heat waves was the greatest in women for overall CVD and respiratory outcomes than in men. In addition, the results stratified by cause-specific mortality indicated that men presented statistically significant higher cumulative RRs for cerebrovascular diseases and ischemic stroke than women in several heat waves and cold spell definitions. In contrast, the higher risk of mortality among women than in men was due to ischemic heart disease on cold events and only a few cases of ischemic heart disease and COPD during heat events. We did not find any statistically significant results for ischemic stroke in women on extremely cold and hot days.

During heat wave events, no higher RRs on added effects compared to the overall effects were detected in men on heat wave events, and we only observed a small increase in added effects in COPD outcomes in women for all temperature thresholds with $\geq 2$ consecutive days. In addition, the results showed only a few higher added effects in women (CVD, respiratory disease, hemorrhagic stroke, and ischemic heart diseases) and men (respiratory, cerebrovascular diseases, and ischemic stroke) in some cold spell definitions.

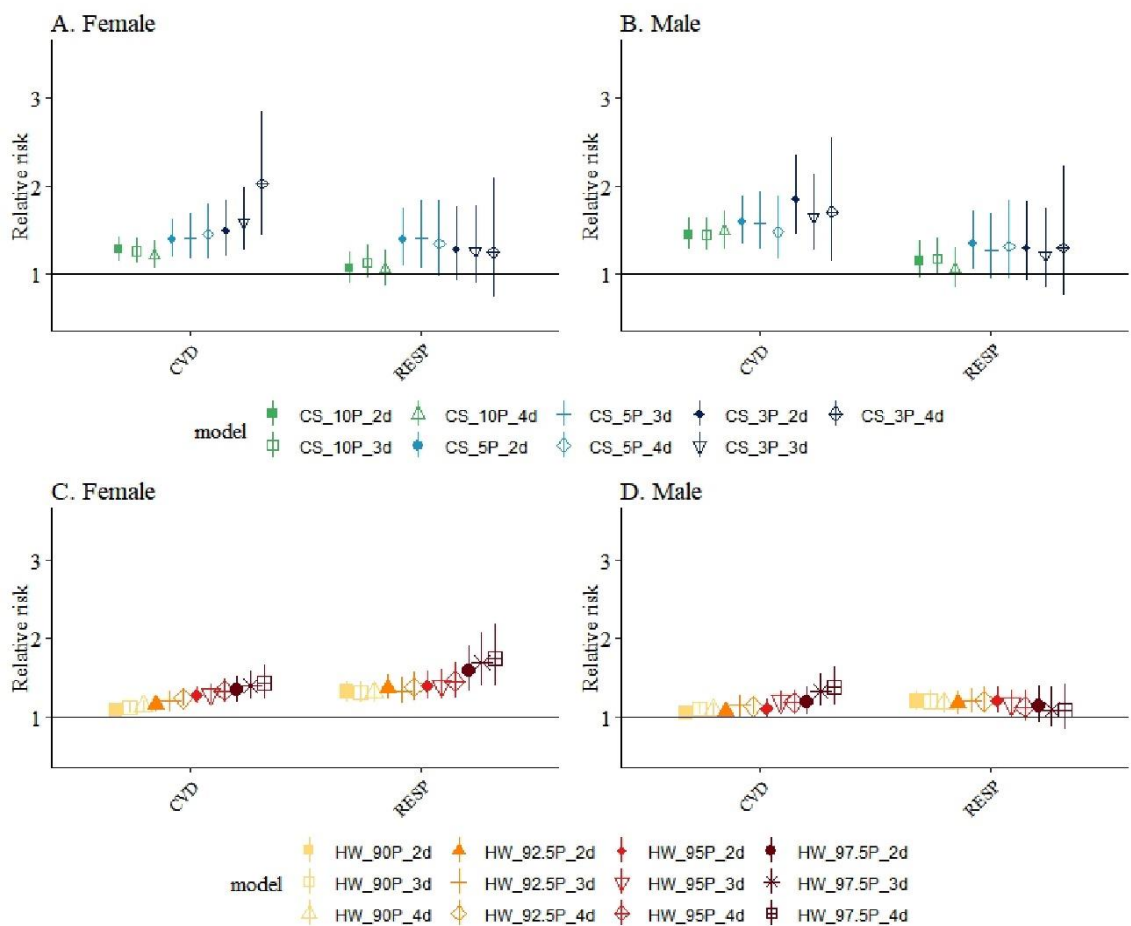

Figure 4: Effects of cold spells (A and B) and heat waves (C and D) on cardiovascular diseases (CVD) and respiratory diseases (RESP) mortality stratified by sex (women and men)

**4. Discussion**

This study assessed the risk of death among older adults (aged ≥65 years) due to CVD, respiratory diseases, cerebrovascular diseases, ischemic stroke, hemorrhagic stroke, ischemic heart disease, and COPD during the occurrence of ETEs from 2006 to 2015. To the best of our knowledge, this is the first study to quantify the impact of cold spells on mortality in Brazil and the first to estimate the association between heat waves and cold spells and cause-specific mortality in São Paulo under different definitions of ETEs.

Our results showed a significantly high risk of mortality associated with heat wave events in several definitions. The highest risk of death during heat waves was identified mostly in the higher temperature threshold (95$^{th}$ and 97.5$^{th}$ percentile) with a duration of ≥3 and ≥4 consecutive days. Similar findings were found in a multi-community study of 400 communities in 18 countries. According to Guo et al. (2017), the higher temperature thresholds of the heat wave definitions (daily mean temperature, 95$^{th}$ and 97.5$^{th}$ percentile) presented a higher risk of all-cause and non-accidental mortality in most communities, including São Paulo and other cities in Brazil.

We observed an increase in deaths during cold spells compared to those on non-cold-spell days. Furthermore, the extremely cold spell threshold definition (percentile 3) presented the highest significant risk of mortality due to CVD, cerebrovascular diseases, and ischemic and hemorrhagic strokes among older adults. A Chinese study found a higher risk of mortality due to respiratory diseases and COPD during cold spells, and they also observed that older individuals were more vulnerable (Chen et al., 2019).

Older adults are particularly susceptible to extreme high and low temperatures compared to younger adults because of their lower thermoregulatory capacity to thermal variations and individual behaviors (such as social isolation, tobacco use, obesity, comorbidities, and frailty) (Balmain et al., 2018; Kenny et al., 2010; Tansey and Johnson, 2015). Physiological responses to high and low temperature exposure may induce changes in the metabolic rate, blood pressure, blood viscosity, cholesterol levels, cardiac output, thromboembolism, and bronchoconstriction (Keatinge et al., 1986; Koskela et al., 1996; Tansey and Johnson, 2015), which can lead to a higher risk of death and hospital admissions due to ischemic heart diseases, stroke, and COPD during ETEs (Gao et al., 2019; Monteiro et al., 2013; Yang et al., 2019; Yin et al., 2018). Medication use can also contribute to a reduction in thermoregulation capacity (Kenny et al., 2010).

Zhao et al. (2019) observed that the risk of CVD hospital admissions during heat wave days decreased in Brazil. However, our findings showed that ETEs significantly increased cardiovascular and respiratory mortality, which is consistent with the findings of previous studies (Anderson and Bell, 2009; Cheng et al., 2019; Fouillet et al., 2006; Yang et al., 2019). The differences in the risk of mortality and hospital admissions due to CVD may be associated with the heat effects on vulnerable individuals. People with comorbidities are more likely to die before receiving medical assistance or been able to be admitted to a hospital on heat wave events (Zhao et al., 2019).

Nevertheless, there are limited studies on the association between ETEs and cause-specific mortality, particularly the subtypes of stroke, ischemic heart disease, and COPD. This cause-specific death is a public health concern because of the high mortality rates worldwide, which can increase with the occurrence of ETEs. According to the World Health Organization (2021), approximately 3.17 million people die from COPD each year, and >15.2 million deaths occur globally due to stroke and heart attack. Calazans and Queiroz (2020) estimated that Brazil has the most significant impact on adult mortality due to cardiovascular diseases compared to 10 other Latin American countries. Another study reported that ischemic heart disease is the first cause of the years of life lost, among Brazilians, followed by stroke in the fourth, and COPD in the eighth position (Marinho et al., 2018). Therefore, it is extremely important to comprehend the cause-specific burden related to ETEs, and further studies should be conducted.

We found significant results and substantial variations in the RR associations between the overall effects of heat waves and cold spell events on cause-specific mortality in older adults. It was possible to identify that there was a high risk of death for all cerebrovascular diseases and ischemic stroke in almost all definitions of heat waves (except the HW_90P_4d classification of ischemic stroke). Statistically significant results were found only in two cold spell definitions for hemorrhagic stroke mortality (CS_3P_3d and CS_3P_4d), and no significant RR was observed during the occurrence of heat wave events. The ischemic heart disease and COPD results showed an increase in mortality only in a few definitions of heat wave events. Yin et al. (2018) found similar results in a heat wave study conducted in China (Yin et al., 2018). Furthermore, they verified a higher risk of mortality due to ischemic stroke than that of hemorrhagic stroke on heat wave days.

In contrast, cold spell-related mortality was identified in all CVD categories and some ETE classifications of cerebrovascular diseases, ischemic stroke, hemorrhagic stroke (CS_10P_4d CS_5P_2d, CS_3P_3d, and CS_3P_4d), ischemic heart disease, and respiratory diseases outcomes. In contrast to the findings of other studies, no significant RR was found for COPD during cold spells (Chen et al., 2019; Han et al., 2017). The intensity and duration of the effects of heat waves and cold spells on mortality have also varied considerably in other studies (Chen et al., 2019; Yang et al., 2019).

Since few studies have investigated heat wave- and cold spell-related cause-specific mortality, especially stratified by sex, our study can provide substantially contributing data on this matter. We identified a significantly higher mortality risk in women than in men on heat waves days for overall cardiovascular and respiratory diseases. Recent studies have shown that women are more vulnerable to heat effects than men (Fouillet et al., 2006; van Steen et al., 2019; Yang et al., 2019). Nonetheless, the responses to extreme heat and cold event exposure in women and men are inconsistent in the literature and require further investigation.

Epidemiological investigations indicate that women may be more affected by high temperatures than men because of the changes in reproductive hormones, higher average life expectancy, and other physiological and thermoregulatory responses to heat stress (van Steen et al., 2019). In contrast, men could be at a higher risk of mortality due to some health outcomes during extreme temperature exposure because they have more cardiovascular diseases and unhealthy behaviors, and are less likely to undergo regular check-ups and seek health care for pre-existing health conditions than women (Crimmins et al., 2019; Rogers et al., 2010).

Our findings suggest a clear distinction in the risk of death stratified by cause-specific mortality and that stratified by sex. We verified that men are more vulnerable to die from ischemic stroke during heat waves and cold spells than women, while women presented a higher risk of dying from ischemic heart diseases during a cold spell and COPD in a few heat wave definitions than men. There was no statistically significant RR for ischemic stroke among women during heat waves and cold spells.

This study also investigated the added effects on mortality in older adults. We identified significant added effects of ETEs on cause-specific mortality and stratified them by sex. Compared to the overall effects models of ETEs, a higher RR of added effects was observed only in a few cold spell event definitions for the health outcomes and stratified by sex. We did not find an increase in the added effects on heat wave days, except for a small increase in the added effects in women due to COPD for all temperature thresholds with ≥2 consecutive days compared with the overall model results. The added effects on mortality remain inconclusive and controversial. Many studies (Gasparrini and Armstrong, 2011; Guo et al., 2017; Lee et al., 2018) have found significant results decomposing the main (independent effects of daily high temperature) and added effects on heat waves or cold spell days, while others have chosen modeling approaches without temperature adjustment or have reported inconsistent results of the added effects (Chen et al., 2019; Yang et al., 2019; Zhao et al., 2019).

In addition, our findings that the overall and added effects are higher on cold spell days than heat wave days can be partly explained by the hypothesis of the acclimatization of the population (Anderson and Bell, 2009). People living in warmer environmental conditions are more adapted to extreme heat events and are more vulnerable to extreme cold events. Guo et al. (2017) showed that heat wave-related mortality was higher in moderately hot and cold areas than in hot and cold areas.

Moreover, the effects of ETEs on mortality in older adults found in this study may also be related to the combination of physiological factors, individual behaviors, the urban environment, and socioeconomic status. Poor household conditions interfere directly and indirectly with thermal comfort (e.g., household quality, thermal building capacity, and lack of a heating and air conditioning system); lower levels of education, low income, and impact of the urban heat island intensity, especially in urbanized areas, are also potential risk factors for health outcomes (Almendra et al., 2017; Gao et al., 2019; Heaviside et al., 2016; Sera et al., 2019; Tan et al., 2010). In addition, future investigations are needed to better understand the higher risk of death during ETEs, particularly in the city of São Paulo, where about 2 million people live in slums (*favelas*) or under inadequate household and sanitation conditions (IBGE, 2020).

Further investigations should be conducted for the development of an ETE warning system in the city of São Paulo and planning measures to reduce the impact of heat waves and cold spells on the population health.

Finally, some limitations of this study need to be noted. We only obtained the temperature and humidity data from one meteorological station to represent the entire city. Furthermore, we did not address the association between the mortality burden during heat waves and cold spells with other risk factors, such as the built environment and socioeconomic status.

## 5. Conclusions

This study highlights the impact of heat waves and cold spell events on cause-specific mortality among older adults stratified by sex. Our findings allow us to advance the understanding of the relationship between climate and health in the city of São Paulo.

The results provide substantial evidence for public health managers and urban planners to implement and promote preventive measures to reduce the impact of heat wave and cold spell events. Development of strategies, such as the heat relief network with emergency cooling centers by the Canadian government during heat waves and expansion of shelters for the most vulnerable groups during cold events, are fundamental to mitigate the negative impacts on health outcomes. In addition, warning can be provided through the news and community health agents to educate and inform the population regarding the effects of ETEs, especially for people aged ≥65 years. Health units linked to the Brazilian Unified Health System as well as the private and supplementary systems must be prepared for immediate assistance in cases of hospital emergency due to myocardial infarction and stroke, for which prompt assistance can prevent death or sequelae.

Furthermore, improvements in the urban microclimate can be achieved through interventions in the built environment. The identification of urban heat islands can also allow urban planners to mitigate them by implementing green areas at strategic places of the city.

**Author contributions**

**Sara Lopes de Moraes** – Conceptualization; Investigation; Data curation; Formal analysis; Methodology; Software; Writing - original draft; Writing - review & editing.

**Ricardo Almendra** – Investigation; Methodology; Supervision; Validation; Visualization, Writing - original draft; Writing – review & editing.

**Ligia Vizeu Barrozo** – Conceptualization; Investigation; Project administration; Supervision; Validation; Visualization; Writing – original draft; Writing - review & editing

**Acknowledgements**

This work was supported by São Paulo Research Foundation (FAPESP) [grant number 2018/25462-0] and the Coordenação de Aperfeiçoamento de Pessoal de Nível Superior (CAPES) – Financial code 001 and the. RA received support from the Centre of Studies in Geography and Spatial Planning (CEGOT), funded by national funds through the Foundation for Science and Technology (FCT) under the reference UIDB/04084/2020.

**Funding**

This work was supported by São Paulo Research Foundation (FAPESP) [grant number 2018/25462-0] and the Coordenação de Aperfeiçoamento de Pessoal de Nível Superior (CAPES) – Financial code 001.

27Sera, F., Armstrong, B., Tobias, A., Vicedo-Cabrera, A.M., Åström, C., Bell, M.L., Chen, B.Y., De Sousa Zanotti Stagliorio Coelho, M., Correa, P.M., Cruz, J.C., Dang, T.N., Hurtado-Diaz, M., Do Van, D., Forsberg, B., Guo, Y.L., Guo, Y., Hashizume, M., Honda, Y., Iñiguez, C., Jaakkola, J.J.K., Kan, H., Kim, H., Lavigne, E., Michelozzi, P., Ortega, N.V., Osorio, S., Pascal, M., Ragettli, M.S., Ryti, N.R.I., Saldiva, P.H.N., Schwartz, J., Scortichini, M., Seposo, X., Tong, S., Zanobetti, A., Gasparrini, A., 2019. How urban characteristics affect vulnerability to heat and cold: A multi-country analysis. Int. J. Epidemiol. 48, 1101–1112. https://doi.org/10.1093/ije/dyz008

Song, X., Wang, S., Li, T., Tian, J., Ding, G., Wang, Jiaxin, Wang, Jiexin, Shang, K., 2018. The impact of heat waves and cold spells on respiratory emergency department visits in Beijing, China. Sci. Total Environ. 615, 1499–1505. https://doi.org/10.1016/j.scitotenv.2017.09.108

Tan, J., Zheng, Y., Tang, X., Guo, C., Li, L., Song, G., Zhen, X., Yuan, D., Kalkstein, A.J., Li, F., Chen, H., 2010. The urban heat island and its impact on heat waves and human health in Shanghai. Int. J. Biometeorol. 54, 75–84. https://doi.org/10.1007/s00484-009-0256-x

Tansey, E.A., Johnson, C.D., 2015. Recent advances in thermoregulation. Adv. Physiol. Educ. 39, 139–148. https://doi.org/10.1152/advan.00126.2014

Tong, S., Wang, X.Y., Yu, W., Chen, D., Wang, X., 2013. The impact of heatwaves on mortality in Australia: A multicity study. BMJ Open 4, 1–6. https://doi.org/10.1136/bmjopen-2013-003579

van Steen, Y., Ntarladima, A.M., Grobbee, R., Karssenberg, D., Vaartjes, I., 2019. Sex differences in mortality after heat waves: are elderly women at higher risk? Int. Arch. Occup. Environ. Health 92, 37–48. https://doi.org/10.1007/s00420-018-1360-1